# Graphene Hypersurface for Manipulation of THz Waves

Sasmita Dash[1,a], Christos Liaskos[2,b], Ian F. Akyildiz[3,c] and Andreas Pitsillides[1,d]

[1]Computer Science Department, University of Cyprus, Cyprus
[2]Foundation for Research and Technology - Hellas (FORTH), Greece
[3]School of Electrical and Computer Engineering, Georgia Institute of Technology, USA

[a]sdash@ec.iitr.ac.in, [b]cliaskos@ics.forth.gr, [c]ian@ece.gatech.edu, [d]andreas.pitsillides@ucy.ac.cy

**Keywords:** Graphene, Metasurface, THz wave, Absorption, Reconfiguration, Anomalous reflection.

**Abstract.** In this work, we investigated graphene hypersurface (HSF) for the manipulation of THz waves. The graphene HSF structure is consists of a periodic array of graphene unit cells deposited on silicon substrate and terminated by a metallic ground plane. The performance of the proposed HSF is numerically analyzed. Electromagnetic parameters of HSF such as permeability, permittivity, and impedance are studied. The proposed graphene HSF has active control over absorption, reflection, and transmission of THz waves. The graphene HSF provides perfect absorption, zero reflection and zero transmission at resonance. Moreover, the graphene HSF structure has the advantage of anomalous reflection and frequency reconfiguration. Incident waves can be reflected in the desired direction, depending on the phase gradient of the HSF and the perfect absorption is maintained at all reconfigurable frequencies upon reconfiguration. The results reveal the effectiveness of the graphene HSF for the manipulation of THz waves.

**Introduction**

Terahertz (THz) wave, frequency ranging from 0.1 to 10 THz, lies in between microwave and infrared frequencies in the electromagnetic spectrum. In the last decade, THz technology gained huge research attention due to promising applications in defense, medical, earth and space science, material characterization, communication, sensing and imaging [1]. Owing to its two-dimensional nature and the unique properties at THz frequencies, graphene has attracted a lot of attention to THz research community. The most important properties of graphene at THz frequency is the ability to support the propagation of the surface plasmon polariton (SPP) waves [2]. The graphene plasmons possess more confinement, low loss, and good tunability. Furthermore, the conductivity of graphene is determined by the Fermi level and can be tuned by the application of electric field, and/or chemical doping. Due to the properties of conductivity function and plasmonic effect at THz frequencies, graphene-based devices enable high miniaturization and reconfiguration [3], [4]. The unique property makes graphene a promising material for various THz applications such as antennas [3-5], absorbers [6], biosensors [7], modulator [8], and photodetector [9].

In the past decade, metamaterials (MM) have attracted more research attention from physics, material science, and engineering community. Recently, metasurfaces (MSF), a two-dimensional MM, have emerged at the frontier of metamaterial research. MM and MSF have gained great attention due to its ability to control the wave propagation by patterning planar subwavelength structure. The unique feature of MM and MSF is that the unit cell parameters can be controlled to acquire the desired electromagnetic properties. Hypersurface (HSF), a programmable or intelligent MSF, design exhibits a high degree of MSF pattern customization and allows for a high degree of tunability [10]. HSF empowered wireless environment enables the optimization of the propagation factor between wireless devices [10]. There has recently been a keen interest in the use of graphene to control the THz waves. The strong optical response arising from the graphene surface plasmons enables novel MM [11]. Because of its unique properties at THz, graphene can be a promising candidate for THz HSF.

In this work, we have presented a graphene HSF structure, which manipulates incoming THz wave. This work investigates the design and analysis of the graphene HSF structure. Electromagnetic parameters permeability, permittivity, and impedance are studied. The performance of reflection, absorption, transmission, anomalous reflection and reconfiguration of proposed graphene THz HSF structure is analyzed.

**Characteristics of Graphene**

Graphene is a single layer of carbon atoms arranged in a honeycomb lattice form. The unique electronic, optical, mechanical and thermal properties of graphene at THz frequency such as electron mobility of 2, 00,000 $cm^2V^{-1}s^{-1}$, 97.7 % of transparency, thermal conductivity of 5,000 $Wm^{-1}K^{-1}$, tensile strength of 1.5 TPa, breaking strength of 42 N/m, *etc.*, opened plethora of applications in numerous fields [12]. The conductivity of graphene is described by the Kubo formalism [13], which is the contribution from both intraband transition and interband transition.

$$\sigma_g = \frac{je^2(\omega - j\tau^{-1})}{\pi\hbar^2} \times \left[ \frac{1}{(\omega - j\tau^{-1})^2} \int_0^\infty \varepsilon \left( \frac{\partial f_d(\varepsilon)}{\partial \varepsilon} - \frac{\partial f_d(-\varepsilon)}{\partial \varepsilon} \right) d\varepsilon - \int_0^\infty \frac{f_d(-\varepsilon) - f_d(\varepsilon)}{(\omega - j\tau^{-1})^2 - 4(\varepsilon/\hbar)^2} d\varepsilon \right] \quad (1)$$

where $e$ is the electronic charge, $k_B$ is the Boltzmann's constant, $\hbar$ is the reduced Planck's constant, $T$ is the temperature, $\mu_c$ is the chemical potential, $\omega$ is the angular frequency, scattering rate $\Gamma = 1/2\tau$, $\tau$ is the relaxation time related to impurities and defects in graphene, $f_d(\varepsilon) = \left[ e^{(\varepsilon - \mu_c)/\kappa_B T} + 1 \right]^{-1}$ is the Fermi-Dirac distribution. In THz frequency regime, the conductivity of graphene is governed by intraband contribution, which can be approximated as

$$\sigma_s(\omega, \mu_c, \tau, T) = -j \frac{e^2 \kappa_B T}{\pi\hbar^2(\omega - j\tau^{-1})} \left[ \frac{\mu_c}{\kappa_B T} + 2\ln\left(e^{-\mu_c/\kappa_B T} + 1\right) \right] \quad (2)$$

The complex graphene surface conductivity of graphene can be adjusted by controlling the chemical potential $\mu_c$. The chemical potential can be dynamically adjusted using electric field effect via DC bias voltage or chemical doping, thereby tuning the graphene conductivity. An applied electric field bias injects more electron or hole carriers, which allows the dynamic control of both real and imaginary parts of the conductivity. The applied electric field $E_0$ can be approximated as [14].

$$E_0 = \frac{q_e}{\pi\varepsilon_0 \hbar^2 v_f^2} \int_0^\infty \varepsilon \left[ f_d(\varepsilon) - f_d(\varepsilon + 2\mu_c) \right] d\varepsilon \quad (3)$$

The permittivity of graphene can be obtained by its surface conductivity as [15]

$$\varepsilon_g = 1 + i\frac{\sigma_g}{t_g \varepsilon_0 \omega} \quad (4)$$

where $t_g$ is the thickness of the graphene, and $\varepsilon_0$ is the permittivity of vacuum.
The above equations indicate that the electromagnetic properties of graphene can be dynamically controlled by DC bias voltage or chemical doping; leading to that the electromagnetic characteristics of the graphene-based structure can also be dynamically controlled.

**Design and Analysis of Graphene HSF and Unit Cell**

A smart HSF using graphene unit cell at THz band is presented in this work. The graphene HSF structure is consists of a two-dimensional periodic array of graphene elements. Fig. 1 (a) and (b) show the proposed graphene HSF structure and its unit cell respectively. The graphene patch is deposited on silicon substrate of 9 μm of height and terminated by a metal ground plane. An external biasing DC voltage is applied between the graphene layer and a polycrystalline silicon layer of 50 nm thickness for adjusting the chemical potential of graphene. A 50 nm thickness of the $SiO_2$ layer is inserted in between the graphene and polycrystalline silicon layer as a spacer.

The periodic array consists of graphene unit cells with periodicity $P$ (=14μm). The graphene unit cell has the dimension of $d \times d$, set along the x- and y-directions. The length of the graphene unit cell is $d$ (=8.5 μm). The frequency is $f$ = 2.5 THz. Due to plasmonic wave propagation, the plasmonic resonance occurs at very small sizes $\lambda/14$ *i. e.,* below $\lambda/10$. To achieve a thin structure, the thickness of the substrate is considered as $h = \lambda/13$ *i. e.,* below $\lambda/10$. Graphene is modeled as a thin resistive sheet using surface conductivity condition (Eq. 2). The relaxation time of graphene $\tau = 1$ pS and room temperature $T = 300K$ is considered in this work. The chemical potential is controlled from 0.5 to 0.65 eV by applying external gate voltage.

The proposed graphene unit cell structure and HSF structure are validated in the CST simulator. For the simulation of the graphene unit cell, a Floquet port is applied to excite the impinging plane wave onto the unit cell.

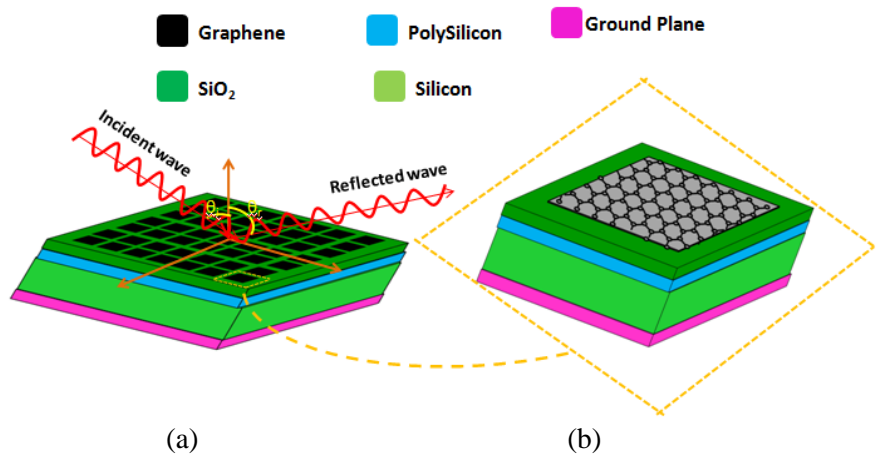

(a)          (b)

Figure 1. Schematic of (a) Graphene HSF, and its (b) unit cell.

Permittivity and permeability are constitutive electromagnetic parameters of materials. In order to understand the graphene HSF further, permittivity and permeability of HSF are also investigated. The real and imaginary parts of permittivity and permeability are shown in Figure 2 (a) and 2(b) respectively. From Figures, It can be marked that the real part of permittivity and permeability are negative. For perfect absorption, the complex permittivity and permeability of the proposed graphene HSF structure must exhibit electric and magnetic resonances, respectively, which are defined by the Lorentz model. It is widely known that artificial MM and MSF show Lorentz dispersion. The refractive index of MSF/HSF will have a negative value when the real parts of constitutive parameters permittivity $\varepsilon_r$ and permeability $\mu_r$ are simultaneously negative; otherwise, the refractive index will have a positive value. Constitutive electromagnetic parameters values at resonance are listed in Table 1.

Table 1. Constitutive electromagnetic parameters value at resonance.

| Parameters | Real Permittivity $\varepsilon_r$ | Imaginary permittivity $\varepsilon_i$ | Real Permeability $\mu_r$ | Imaginary Permeability $\mu_i$ | Real Impedance $Z_r$ | Imaginary Impedance $Z_i$ |
|---|---|---|---|---|---|---|
| **Value** | -2 | 13 | -20 | 20 | 1 | 0 |

Further, to study the input impedance characteristic of the proposed structure, we obtained the real part and the imaginary part of the normalized input impedance of the structure. Fig. 2 (c) shows the real and the imaginary part of the normalized input impedance as a function of frequency for the proposed graphene HSF structure. From Fig. 2 (c), it can be observed that the imaginary part of the normalized input impedance is zero at central frequency 2.5 THz. The real part of the normalized input impedance at the central frequency 2.5 THz is unity, i.e. $Z_{in}/Z_0 = 1$, where $Z_{in}$ and $Z_0$ are input impedance and free space impedance respectively. Therefore, the impedance matching condition for perfect absorption is satisfied with the proposed graphene HSF structure.

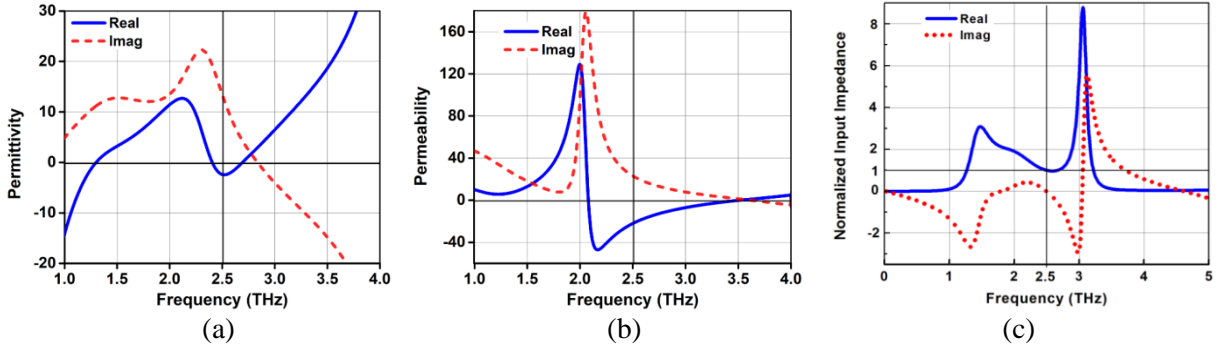

Figure 2. The real and imaginary part of (a) Permittivity, (b) Permeability and (c) Normalized input impedance.

## Results and Discussion

**Reflection, Absorption, and Transmission.** The proposed graphene HSF provides 100% absorption, zero reflection and zero transmission at resonance 2.5 THz. Reflection, absorption and transmission coefficient of the proposed graphene HSF structure under normal incidence is shown in Fig. 3 (a).

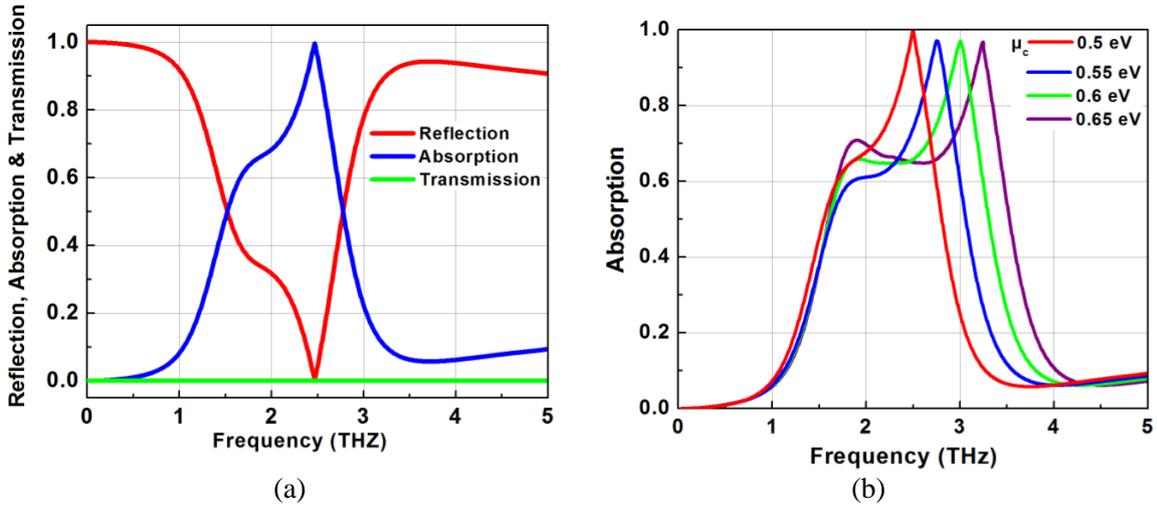

Figure 3. (a) Reflection, transmission, and absorption spectra of graphene HSF, and (b) Frequency reconfiguration of graphene HSF.

**Anomalous Refection.** The spatial phase distribution on graphene HSF forms a desired beam pattern. Its conceptual illustration is shown in Fig, 4. An incident wave can be reflected in the desired direction, depending on the phase gradient of the HSF or length of supercell. According to Snell's law of reflection, the relationship between the incident angle and reflection angle is generalized to

$$\sin(\theta_r) - \sin(\theta_i) = \frac{\lambda}{2\pi n_i} \frac{d\phi}{dx} \quad (5)$$

where $n_i$ is the refractive index of incidence medium. $\theta_i$ and $\theta_r$ are the angle of incidence and angle of reflection, respectively. $d\phi/dx = 2\pi/D$ represents that only length of supercell $D$ (=$N_c d$, where $N_c$ is the number of graphene unit cell) and wavelength of the incident electromagnetic wave $\lambda$ are important in determining the reflected angle. There is a nonlinear relation between the angle of incidence $\theta_i$ and angle of reflection $\theta_r$, clearly different from specular reflection, as shown in Figure 4. Anomalous reflection angles as per our desired direction are listed in Table 2.

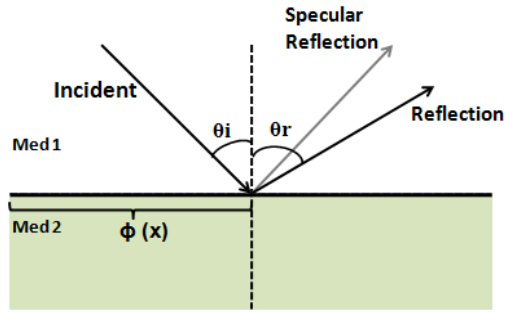

Figure 4. Conceptual illustration of HSF.

Table 2. Anomalous reflection angles.

| Incident angle $\theta_i$ | Number of unit cells $N_c$ | Reflected angle $\theta_r$ |
|---|---|---|
| 15 | 58 | 30 |
|  | 31 | 45 |
|  | 23 | 60 |
|  | 19 | 75 |
| 30 | 68 | 45 |
|  | 38 | 60 |
|  | 32 | 70 |
|  | 29 | 80 |

**Reconfiguration.** The frequency reconfiguration of the graphene HSF is achieved by tuning the chemical potential of graphene. An external biasing DC voltage is applied between the graphene layer and a polycrystalline silicon layer for adjusting the chemical potential of graphene. Fig. 3 (b) shows the impact of tuning of the chemical potential of graphene on the absorption spectra and the frequency. Here, we consider the graphene chemical potential $\mu_c$ over the ranges from 0.5 eV to 0.65 eV. The small change of chemical potential makes a significant shift in the central frequency. The central frequency tuned from 2.5 THz to 3.3 THz as chemical potential $\mu_c$ increases from 0.5 eV to 0.65 eV. Fig. 4 also reveals that perfect absorption performance is maintained at reconfigurable frequencies.

**Conclusion**

In this work, we investigated a graphene HSF for the manipulation of THz waves. The electromagnetic parameters of HSF, such as permeability, permittivity, and impedance are studied. The graphene HSF structure provides perfect absorption without any reflection and transmission. In addition, the performance of anomalous reflection and frequency reconfiguration of proposed graphene THz HSF is studied. The results reveal the effectiveness of the graphene HSF, which can be promising for the manipulation of THz waves.

**Acknowledgments**

This research is supported by the European Union via the Horizon 2020: Future Emerging Topics - Research and Innovation Action call (FETOPEN-RIA), grant EU736876, Project VISORSURF (http://www.visorsurf.eu).